# New Measurements and Phase Shift Analysis of p$^{16}$O Elastic Scattering at Astrophysical Energies

Sergey Dubovichenko[1,2,*], Nassurlla Burtebayev[2,†], Albert Dzhazairov-Kakhramanov[1,2,‡], Denis Zazulin[2], Zhambyl Kerimkulov[2], Marzhan Nassurlla[2], Chingis Omarov[1], Alesya Tkachenko[1], Tatyana Shmygaleva[1,3], Stanislaw Kliczewski[4], Turlan Sadykov[5]

[1]Fesenkov Astrophysical Institute "NCSRT"ASC MID Republic of Kazakhstan (RK)
[2]Institute of Nuclear Physics ME RK Almaty, RK
[3] Kazakh National University al-Farabi Almaty, RK
[4]H. Niewodniczański Institute of Nuclear Physics, Polish Academy of Sciences, Cracow, Poland
[5]Institute of Physics and Technology, Almaty, RK

[‡, 1]*albert-j@yandex.ru*
[*]*dubovichenko@mail.ru*
[†]*burteb@inp.kz*

**Abstract:** The results of new experimental measurements of p$^{16}$O elastic scattering in the energy range of 0.6–1.0 MeV at angles of 40°–160° are given. Phase shift analysis of p$^{16}$O elastic scattering was made using these and other experimental data on differential cross sections in excitation functions and angular distributions at energies of up to 2.5 MeV.

## 1. Introduction

About 20 years ago in our previous work [1], we showed a possible way to describe lithium nuclei in the framework of the potential cluster model (PCM) [2,3]. This PCM takes into account forbidden states (FSs) [3,4] in intercluster potentials that we used in [5]. Finally, the possibility of describing the astrophysical *S*-factors or total cross sections for radiative capture of p$^2$H, n$^2$H, p$^3$H, p$^6$Li, n$^6$Li, p$^7$Li, n$^7$Li, p$^9$Be, n$^9$Be, p$^{10}$B, n$^{10}$B, p$^{11}$B, n$^{11}$B, p$^{12}$C, n$^{12}$C, p$^{13}$C, n$^{13}$C, p$^{14}$C, n$^{14}$C, n$^{15}$N, p$^{15}$N, n$^{14}$N, n$^{16}$O and $^2$H$^4$He, $^3$He$^4$He, $^3$H$^4$He, $^4$He$^{12}$C systems at thermal and astrophysical energies was shown in [2,3,5–15]. Calculations of these 27 processes were made on the basis of modified PCM variant with classification of states according to Young tableaux and forbidden, in some cases, states (MPCM), described in detail in [5,9–16].

Some success of this MPCM can be explained by the fact that intercluster interaction potentials are constructed not only on the basis of known elastic scattering phase shifts but also taking into account classification of cluster states according to Young tableaux [17]. Thus, the elastic scattering phase shifts, extracted from the experimental differential cross sections, taking into account such a classification, allow one to construct interaction potentials of two particles in continuous spectrum.

Continuing study of thermonuclear reactions in the frame of the MPCM with FSs [18] let us consider the $^{16}$O(p,γ)$^{17}$F process, which takes part of the CNO cycle [19] and has additional interest, since it is the reaction at the last nucleus of 1$p$-shell with the forming of $^{17}$F that get out its limit. As we usually assume [16,19], the bound state (BS) of $^{17}$F is caused by the cluster channel of the initial particles, which take part in the reaction.

---
[1] Corresponding author

Many stars, including the Sun, will eventually pass through an evolutionary phase that is referred to as the asymptotic giant branch [20]. This phase involves a hydrogen and a helium shell that burn alternately surrounding an inactive stellar core. The $^{16}$O(p,γ)$^{17}$F reaction rate sensitively influences the $^{17}$O/$^{16}$O isotopic ratio predicted by models of massive (≥4$M_\odot$) AGB stars, where proton captures occur at the base of the convective envelope (hot bottom burning). A fine-tuning of the $^{16}$O(p,γ)$^{17}$F reaction rate may account for the measured anomalous $^{17}$O/$^{16}$O abundance ratio in small grains which are formed by the condensation of the material ejected from the surface of AGB stars via strong stellar winds [21].

Furthermore, these potentials allow one to carry out calculations of some interaction characteristics of the particles involved in the processes of the elastic scattering and reactions. For example, it could be astrophysical $S$-factors of the radiative capture reactions [22] or the total cross sections of these reactions [23]. Therefore, for construction of the potentials of two particles, it is preferable to perform phase shift analysis and to obtain the scattering phases at astrophysical energies, i.e., usually up to 1.0–2.0 MeV. At the same time, all the analyses of p$^{16}$O elastic scattering that have been made so far started from 1.5–2.5 MeV and do not cover the astrophysical energy region per se.

Proceeding to the direct description of the results of our phase shift analysis of the p$^{16}$O elastic scattering at energies of up to 2.0–2.5 MeV, we have already carried out a phase shift analysis of nine systems, namely: n$^3$He, p$^6$Li, n$^{12}$C, p$^{12}$C, $^4$He$^4$He, $^4$He$^{12}$C, p$^{13}$C, p$^{14}$C, and n$^{16}$O, [24,25] mostly at low and astrophysical energies. To perform this analysis, we used data on the differential cross sections in the excitation functions or angular distributions given in the EXFOR database [26] and obtained in the present measurements.

In the case of elastic scattering of nuclear particles with spin 1/2 + 0, the cross section is fully described by two independent spin amplitudes ($A$ and $B$) and can be represented in the form (see, for example, [24]):

$$\frac{d\sigma(\theta)}{d\Omega} = |A(\theta)|^2 + |B(\theta)|^2, \tag{1}$$

where

$$A(\theta) = f_c(\theta) + \frac{1}{2ik}\sum_{L=0}^{\infty}\{(L+1)S_L^+ + LS_L^- - (2L+1)\}\exp(2i\sigma_L)P_L(\cos\theta),$$

$$B(\theta) = \frac{1}{2ik}\sum_{L=0}^{\infty}(S_L^+ - S_L^-)\exp(2i\sigma_L)P_L^1(\cos\theta),$$

$$f_c(\theta) = -\left(\frac{\eta}{2k\sin^2(\theta/2)}\right)\exp\{i\eta\ln[\sin^{-2}(\theta/2)] + 2i\sigma_0\}. \tag{2}$$

Here $S_L^\pm = \eta_L^\pm \exp(2i\delta_L^\pm)$ – scattering matrix, $\delta_L^\pm$ – required scattering phase shifts, $\eta_L^\pm$ – inelasticity parameters, and signs "±" correspond to the total moment of system $J = L \pm 1/2$, $k$ – wave number of the relative motion of particles $k^2 = 2\mu E/\hbar^2$, $\mu$ – reduced mass, $E$ – the energy of interacting particles in the center-of-mass system, $\eta$ – Coulomb parameter.

The multivariate variational problem of finding these parameters at the specified range of values appears when the experimental cross sections of scattering of nuclear particles and the mathematical expressions, which describe these cross sections with certain parameters $\delta_L^J$ – nuclear scattering phase shifts, are known. Using the experimental data of differential cross-sections of elastic scattering, it is possible to find a set of phase shifts $\delta_L^J$, which can reproduce the behavior of these cross-sections with certain accuracy. Quality of description of experimental data on the basis of a certain theoretical function or functional of several variables of Eqs. (1) and (2) can be estimated by the $\chi^2$ method, which is written as

$$\chi^2 = \frac{1}{N}\sum_{i=1}^{N}\left[\frac{\sigma_i^t(\theta) - \sigma_i^e(\theta)}{\Delta\sigma_i^e(\theta)}\right]^2 = \frac{1}{N}\sum_{i=1}^{N}\chi_i^2, \qquad (3)$$

where $\sigma^e$ and $\sigma^t$ are experimental and theoretical, i.e., calculated for some defined values of the scattering phase shifts cross-sections of the elastic scattering of nuclear particles for $i$-angle of scattering, $\Delta\sigma^e$ – the error of experimental cross-sections at these angles, $N$ – the number of measurements. The details of the using by us searching method of scattering phase shifts were given in [27].

## 2. Review of experimental data

One of the first measurements of differential cross sections for p$^{16}$O elastic scattering with phase shift analysis at energies of 2.0–7.6 MeV was made in [28]. This analysis used the results of [29,30] and some other unpublished results in two energy ranges: 2.0–4.26 MeV and 4.25–7.6 MeV. Resonance at 2.66 MeV in the laboratory system (l.s.) for the $^2P_{1/2}$ wave was discussed in detail. Later in [31], polarizations of p$^{16}$O elastic scattering in the region 2.5–5.0 MeV were measured and a new phase shift analysis at these energies was made, which, however, did not show an explicit resonance at 2.66 MeV [32]. Furthermore on the figures in [33] and the table in [34] (referring to [33]) the results of a detailed phase shift analysis of elastic p$^{16}$O scattering is given at energies of 1.5–3.0 and 2.5–3.0 MeV, respectively, and the presence of a narrow resonance at subsequently refined energy 2.663(7) MeV with a width of 19(1) keV was confirmed [33]. This corresponds to the first superthreshold state of $^{17}$F at 3.104 MeV $J^\pi = 1/2^-$ [32] and is matched with the $^2P_{1/2}$ wave in p$^{16}$O elastic scattering.

Processes of p$^{16}$O elastic scattering in the energy range of 1.0–3.5 MeV were considered in many papers (see, e.g., review in [32] and [35,36]). In Particular, in [37,38] the regions of 0.5–0.6 MeV and 2.0–2.5 MeV were examined. In [39] the excitation functions at energies from 0.4 to 2.0 MeV were measured. However, phase shift analysis of the experimental data was never made in any of the studies [35–39]. As a result, currently available phase shift analyses were made in the 1960s and usually begin at 2.0–2.5 MeV and further, at higher energies. There is only one point in the scattering phase shifts at 1.5 MeV, obtained in [33], which has not been confirmed in subsequent studies.

The study of the energy range of 2.0–2.5 MeV to 7–8 MeV and above in the aforementioned works is related to the fact that scattering phase shifts were constructed for further consideration of certain problems in nuclear physics – they did not cover the range of astrophysical energies. We will further consider radiative capture in the field of astrophysical energies from about $10^{-8}$ MeV up to 2.0–2.5 MeV. The results of the aforementioned studies [35–39] and some others on the excitation functions and angular distributions are quite sufficient for performing phase shift analysis and further constructing the potentials for the p$^{16}$O interaction according to the scattering phase shifts.

For this purpose, let us carry out the phase shift analysis of the available experimental data from 0.4–2.5 MeV and obtain the exact form of the scattering phases in this energy region. In addition, we will verify the results of some other phase shift analyses made in the 1960s. Additionally we give here new experimental data obtained in the INP (Almaty), and we will undertake their phase shift analysis. For the energy region from 0.6–1.0 MeV in the angular range from 40°–160°, both the angular distributions at three energies and excitation functions were measured in this experiment to an accuracy of about 5%.

The calculation methods of the differential cross sections that were used in phase shift analysis are well known and described, for example, in a classic work [27]; methods of this analysis and certain previous results are given in [4,16] and in the aforementioned papers [24,25]. In our present analysis, we used the exact values of masses of particles equal to $M_P = 1.00727646577$ atomic mass units (amu) and $M_{16O} = 15.994915$ amu – these values are taken from databases [40,41], respectively. Note that there is no principal importance whether to use the whole or exact values of masses of particles, since the error in the cross sections is typically 5–10%, it will be seen furthermore from the used experimental data. However, in all our calculations of radiative capture processes [2,16], we always use the exact values of masses of particles, as they significantly affect the binding energy. Constant $\hbar^2/m_0$ was assumed to be equal to 41.4686 MeV fm$^2$, where $m_0$ – amu.

Let us first consider the results obtained in the phase shift analysis, which will be made using angular distributions from [33] in the energy range 1.5–3.0 MeV at 4 scattering energies in the angular range 20°–160°. In other words, we repeat the analysis made in [33] in the 1960s. The results of description of the cross sections with phase shifts extracted in our analysis and the phase shifts themselves are shown by the open squares in Fig. 1 compared to the data given in [28,31,33,34].

From these results it is clear that only at the energy of 2.978 MeV does the phase shift from [33], and that obtained here differ by 1.5–2°. For other three energies, the coincidence is less than 1°, and for 2.48 MeV the results are the same. As we shall consider further the proton radiative capture on $^{16}$O at energies up of to 2.5 MeV without taking into account a narrow resonance at 2.66 MeV [32], the region of this resonance, previously studied in phase shift analysis [31] and shown in Fig. 1 by open circles, as well as in the analysis of [28], will not be considered in detail here.

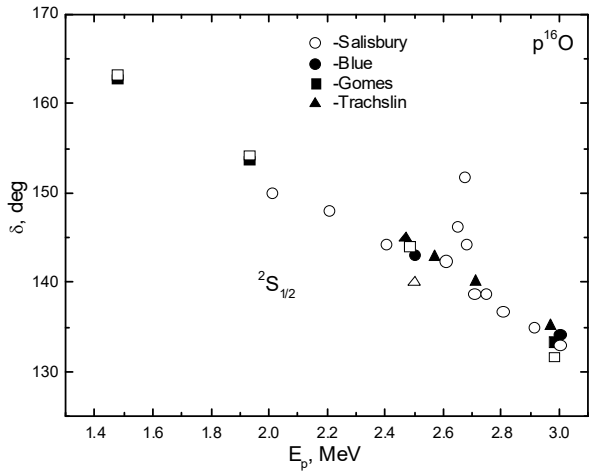 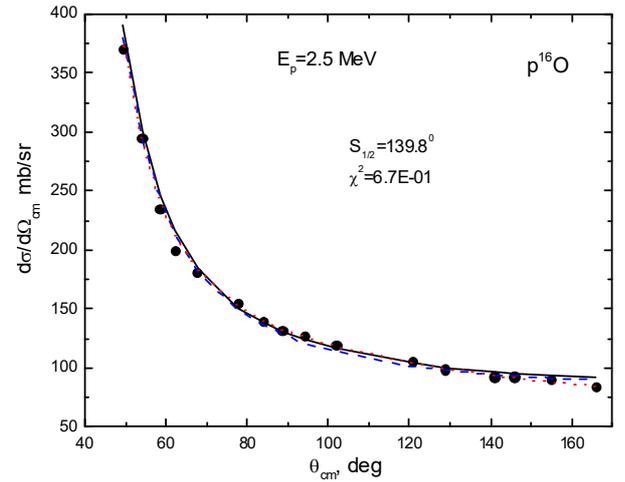

Fig. 1. Phase shifts of p$^{16}$O elastic scattering that we obtained from the angular distributions of [33] - open squares and [31] - open triangle. The remaining notation = data of [28,31,33,34].

Fig. 2. Angular distributions of p$^{16}$O elastic scattering measured in [31]. Different curves – calculation of these cross sections on the basis of different phase shift analyses. Dotted curve – red, dashed curve – blue.

Here is another result for angular distributions of [31] at the energy of 2.5 MeV. Fig. 2 shows the points of differential cross sections measured in the angular distributions, and the solid curve shows the results of our computations of these cross sections with the found phase shifts. The $\chi^2$ value is equal to 0.67 with 10% experimental error, only taking into account a single scattering phase shift, as shown in Fig. 2 and equal to $S_{1/2} = 139.8°$; this phase shift is shown in Fig. 1 by an open triangle. If we add the $P_{1/2}$ scattering phase shift to the analysis, then with 10 iterations [16] we will obtain $\chi^2 = 0.58$ and phase shifts: $S_{1/2} = 140.3°$ and $P_{1/2} = 5.5°$. Ten iterations are needed for converging of $\chi^2$ with an accuracy about 1% (see [4,16]). If we consider also the $P_{3/2}$ phase shift, with the same number of iterations we find $\chi^2 = 0.57$ and phase shifts: $S_{1/2} = 139.7°$, $P_{1/2} = -4.5°$ and $P_{3/2} = 4.6°$. Hence it is clear that taking into account the $P$ scattering phase shifts at an energy that is adjacent to the region of a narrow resonance practically does not change the value of the $S$ phase and does not significantly improve the $\chi^2$ value.

Phase shift analysis has also been made in [31] and following phase shifts were obtained for this energy: $S_{1/2} = 143.2°$, $P_{1/2} = 2.0°$, $P_{3/2} = 2.2°$, $D_{3/2} = 3.2°$ and $D_{5/2} = -1.6°$; the $\chi^2$ value is not given. With these phase shifts in our calculations we obtained the value $\chi^2 = 0.62$ with 10% experimental errors, and the results for the elastic scattering cross sections are given in Fig. 2 by the dashed curve. When we ran a variation of phases given in [31], under our program with 10 iterations, we obtain $\chi^2 = 0.57$ with phases: $S_{1/2} = 140.8°$, $P_{1/2} = -2.8°$, $P_{3/2} = 4.6°$, $D_{3/2} = 3.1°$ and $D_{5/2} = -2.3°$. The scattering cross sections with such phases are given in Fig. 2 by the dotted curve. From this figure and the results of [31], it is clear that taking into account $D$ phase shift does not change the value of $\chi^2$, but the values of $P$ scattering phase shifts change slightly.

## 3. New results for excitation functions and angular distributions

New experimental data on elastic scattering of protons by nuclei oxygen at low energies were measured on electrostatic tandem accelerator UKP-2-1 of Nuclear Physics Institute ME RK [42]. Protons were accelerated to energies $E_p$ = 600–1040 keV. The value of the beam current was limited by the stability of the target and load characteristics of the electronic apparatus and was ranging from 1–80 nA. Calibration of proton energies in the beam was made according to reactions with narrow, well-separated resonances [43]. For this purpose we used $^{27}$Al(p,γ)$^{28}$Si reaction at $E_{p,lab.}$ = 632, 773, 992, 1089 keV and $^{19}$F(p,αγ)$^{16}$O at $E_p$ = 340 keV. The accuracy of beam calibration was equal to ±1 keV. The energy spread of the beam was determined by the width of the front of $^{27}$Al(p,γ)$^{28}$Si reaction yield curve near resonance at $E_p$ = 992 keV (resonance width < 0.1 keV) and did not exceed 1.2 keV [44].

The proton beam passed through a collimation system (i.e. two collimators with diameters of 1.5 mm placed 420 mm apart) and was formed on the target (located at a distance of 100 mm from the last collimator) into a spot with diameter 2 mm. In order to minimize the number of protons scattered from the end faces of the collimators, the thickness of the front wall near the holes was brought to 0.1 mm. A Faraday cup (i.e. a tube with a diameter of 15 mm and a length of 150 mm), located at a distance of 120 mm from the target, was connected to a current integrator, which sent a digital pulse to a scaler, once it collected a portion of charge (0.1 nC or 10 nC). The accumulated charge was determined with an error of not more than 1.5%. To minimize the carbon laydown on a target during the measurements, we used a pumping system consisting of ion and turbomolecular pumps, and inside the scattering chamber a nitrogen traps system was installed. A typical pressure in the chamber was 1.5 x 10$^{-6}$ torr.

In order to detect the scattered protons we used a surface-barrier charged particles detector (the diameter of the bounding diaphragm before the detector was 2 mm; the sensitive area thickness was 0.2 mm). The detector was placed at a distance of 240 mm from the target and was able to move in an angular range from 10°–170°. The error in determining the angle of the detector location did not exceed ± 0.2°. The detector was equipped with the protective tube, which, for all its positions, excluded recording of the protons scattered from the end face of the last collimator and from the Faraday cup. A second similar detector was placed at an angle of 160° relative to the incident beam and was used to monitor the stability of the target. The energy resolution of detectors was equal to 15 keV. A detailed description of the experimental setup for the study of the processes with the charged particles produce in the UKP-2-1 can be found in [45] and references therein.

An aluminum oxide film (Al$_2$O$_3$), used as a target was made using the electrolytic method. Proton energy losses (for an incident proton energy of $E_{p,lab.}$ = 992 keV) after passing the target (Al$_2$O$_3$) were determined by width at half-height of the yield curve of $^{27}$Al(p,γ)$^{28}$Si reaction near resonance at $E_{p,lab.}$ = 992 keV (the target was placed exactly perpendicular to the incident beam) and were found to be 5.4±1.2 keV, which corresponds to the thickness of the target 28±6 μg/cm$^2$ [46]. Such a target thickness satisfied the requirements of mechanical and thermal strength, and at the same time, practically did not affect the spectral line broadening, except for spectral lines obtained at θ$_{c.m.}$ = 72.4°, 92.6°, 103° at $E_{p,lab.}$ = 600 keV, where broadening was due to the target thickness being equal to the broadening due to the detector energy resolution.

Signals from the detectors were amplified and transmitted to two 2024-channel analyzers. The electronics dead time did not exceed 3%. At each proton energy value, the ratio of the area of the peak from the stationary detector because of $^{16}$O(p,p)$^{16}$O and $^{27}$Al(p,p)$^{27}$Al scattering to the reading of the integrator counter was a constant within 4% for all positions of the movable detector. The laboratory energy given in this work corresponds to laboratory proton energy in the center of the target thickness.

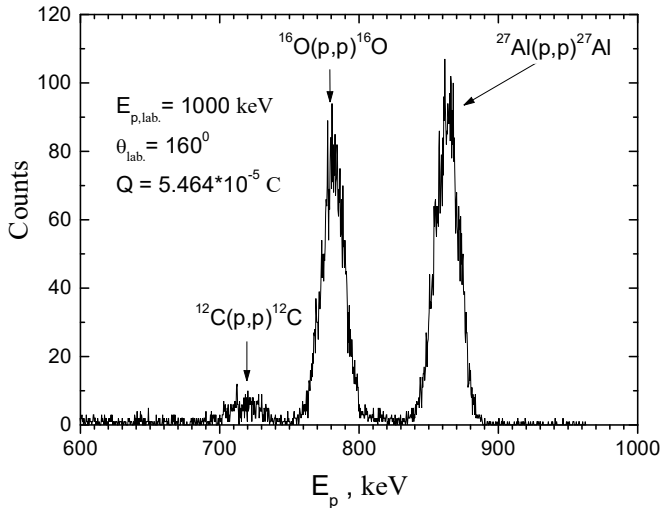

Fig. 3a. Energy spectrum of protons elastically scattered from target nuclei

An example of proton elastic scattering from target nuclei spectrum obtained at $E_{p,lab.} = 1000$ keV is given in Fig. 3a. The peaks from the elastic scattering of protons from $^{12}C$, $^{16}O$ and $^{27}Al$ nuclei are clearly seen in the figure. The presence of a peak from $^{12}C(p,p)^{12}C$ process in the spectrum is due to the carbon laydown on the target surface. We can see also from Fig. 3a that the background of the spectrum is completely determined by the presence of peaks of $^{12}C(p,p)^{12}C$ and $^{27}Al(p,p)^{27}Al$ processes.

The angular distributions of $^{16}O(p,p)^{16}O$ were measured at incident protons energies $E_{p,lab.} = 600$ keV, 800 keV and 1000 keV at angles $\theta_{c.m.} = 41.3°, 62.1°, 72.4°, 92.6°, 103°, 122°, 141°, 151°$, and $160°$. Excitation functions of the $^{16}O(p,p)^{16}O$ were measured in the energy range of $E_{p,lab.} = 600$–1040 keV with a step of 20 keV for two angles 94° and 160° in the center-of-mass system. The target was installed perpendicular to the incident beam for detector positions at angles $\theta_{c.m.} = 41.3°, 62.1°, 122°, 141°, 151°$ and $160°$, and for detector positions at $\theta_{c.m.} = 72.4°, 92.6°$, and $103°$ – at an angle of 45°.

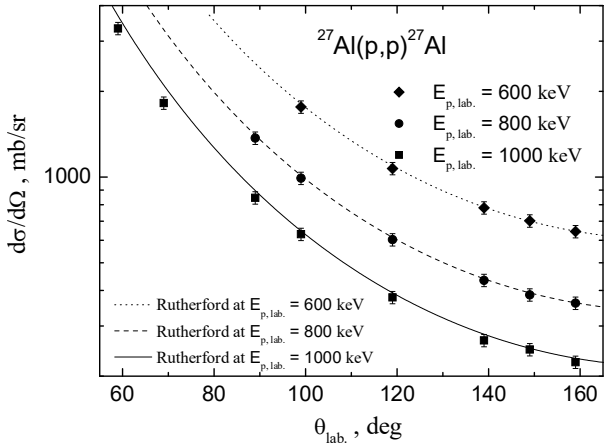

Fig. 3b. The differential cross section of the elastic scattering of protons on $^{27}Al$ with errors of 4%. Symbols are the experimental data of present work; curves = calculations using the Rutherford formula.

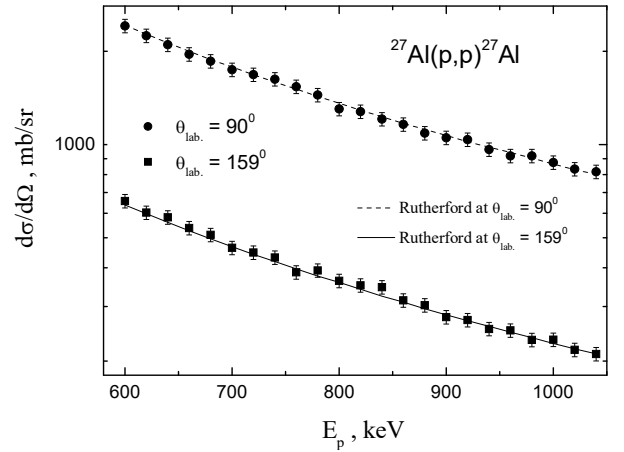

Fig. 3c. Excitation functions of the elastic scattering of protons on $^{27}Al$ with errors of 4%. Symbols are the experimental data of present work; curves = calculations using the Rutherford formula.

The amount of counts in the spectral peak with the preliminarily subtracted background divided by the integrator counter value was taken as yield of the $^{16}O(p,p)^{16}O$ elastic scattering. Statistical error in the determination of the yields (including errors introduced by background subtracted) was less than 3.5% for all positions of the detector and energies of incident protons.

The spectra where peaks from $^{12}C(p,p)^{12}C$, $^{16}O(p,p)^{16}O$ and $^{27}Al(p,p)^{27}Al$ processes overlapped were analyzed using information about the differential cross sections of

$^{12}$C(p,p)$^{12}$C, taken from [47], and $^{27}$Al(p,p)$^{27}$Al. While the number of $^{12}$C nuclei in the target was determined by the spectrum closest to that analyzed, where the peak from $^{12}$C(p,p)$^{12}$C is well separated. For spectra with overlapping peaks, the yield of elastic $^{12}$C(p,p)$^{12}$C scattering does not exceed 10% of the yield of $^{16}$O(p,p)$^{16}$O. Differential cross sections of $^{27}$Al(p,p)$^{27}$Al were assumed as purely Rutherford. This assertion is based on the data shown in Figs. 3b,c, where the differential cross sections (Fig. 3b) and the excitation function (Fig. 3c) for the $^{27}$Al(p,p)$^{27}$Al are given, which are the results of the processing the spectra, where the peaks from $^{27}$Al(p,p)$^{27}$Al scattering are separated reliably (the errors in the determination of the differential cross sections are about 4%). In the cases where peaks from $^{12}$C(p,p)$^{12}$C, $^{16}$O(p,p)$^{16}$O and $^{27}$Al(p,p)$^{27}$Al processes separate completely (Fig. 3a) or overlap weakly, the background for each peak receives the form of a trapezium; for peaks $^{16}$O(p,p)$^{16}$O and $^{27}$Al(p,p)$^{27}$Al it was not greater than 2%. Finally, the differential cross sections of the $^{16}$O(p,p)$^{16}$O were obtained with an error of about 5 and 10% by normalizing the $^{16}$O(p,p)$^{16}$O yields to the normalization factor, which was derived by normalizing the $^{27}$Al(p,p)$^{27}$Al yields to the Rutherford cross sections for $^{27}$Al(p,p)$^{27}$Al. Error of 10% relates to the range $\theta_{c.m.}$ = 41.3°, 62.1° and 72.4° at the energy of $E_{p\,lab.}$ = 600 keV; $\theta_{c.m.}$ = 41.3°, 62.1° and 72.4° at the energy of $E_{p\,lab.}$ = 800 keV; $\theta_{c.m.}$ = 41.3° at the energy of $E_{p\,lab.}$ = 1000 keV, where scattering peaks overlap strongly. The error is determined essentially by ambiguity of normalization coefficient (or target thickness).

Excitation functions and differential cross sections of elastic scattering of protons by $^{16}$O, obtained here are given in Tables 1 and 2, respectively. Within the errors, the results of our experiment coincide with the published data in the overlapping areas. At angles of $\theta_{c.m.}$ = 41.30°, 62.10°, 72.40° and at the energies of $E_{p\,lab.}$ = 600 keV and 800 keV; and at $\theta_{c.m.}$ = 41.3°, 62.1° and at $E_{p,lab.}$ = 1000 keV the experimental cross sections coincide with Rutherford cross sections (to an accuracy of about 10%). However, the ratio of the experimental cross sections to the Rutherford cross sections $\sigma_{exp.}/\sigma_{R.}$ rises monotonically with the increase of angles, reaching at $\theta_{c.m.}$ = 160° and $E_{p,lab.}$ = 1000 keV the value of 1.42 ± 0.07, which is in a good agreement with the literature data and indicates the essential contribution of nuclear forces to the formation of the cross section in this region. This cross section increase at back angles leads to a visible decrease of the scattering phase shift for the partial $^2S_{1/2}$ wave (see below).

Table 1. Excitation function of p$^{16}$O elastic scattering (errors are about 5%).

| | 160°, c.m. | 94°, c.m. |
|---|---|---|
| $E_{p,lab.}$, keV | dσ/dΩ, c.m., (mb/sr) | dσ/dΩ, c.m., (mb/sr) |
| 600 | 318 | 944 |
| 620 | 298 | 890 |
| 640 | 276 | 852 |
| 660 | 258 | 830 |
| 680 | 254 | 754 |
| 700 | 228 | 734 |
| 720 | 226 | 691 |
| 740 | 219 | 642 |
| 760 | 205 | 612 |
| 780 | 202 | 576 |
| 800 | 192 | 559 |
| 820 | 177 | 529 |
| 840 | 188 | 509 |
| 860 | 170 | 488 |
| 880 | 173 | 476 |
| 900 | 165 | 462 |
| 920 | 163 | 431 |
| 940 | 153 | 419 |
| 960 | 147 | 415 |
| 980 | 145 | 404 |
| 1000 | 145 | 394 |
| 1020 | 141 | 373 |
| 1040 | 134 | 372 |

Table 2. Angular distributions of p$^{16}$O elastic scattering.

| θº, c.m. | dσ/dΩ, c.m., (mb/sr) | | |
|---|---|---|---|
| | 600 keV, lab. | 800 keV, lab. | 1000 keV, lab. |
| 41.3 | 17068±10% | 9271±10% | 5208±10% |
| 62.1 | 3712±10% | 1967±10% | 1201±5% |
| 72.4 | 2128±10% | 1195±10% | 814±5% |
| 92.6 | 943±5% | 590±5% | 381±5% |
| 103 | 704±5% | 461±5% | 293±5% |
| 122 | 475±5% | 288±5% | 208±5% |
| 141 | 373±5% | 219±5% | 161±5% |
| 151 | 332±5% | 203±5% | 146±5% |
| 160 | 312±5% | 193±5% | 139±5% |

## 4. New phase shift analysis

As was mentioned in [39] the excitation functions of elastic p$^{16}$O scattering at energies from 0.4–2.0 MeV at 171.5° were measured; however, as far as we know, phase shift analysis of these data was not made. The results of our phase shift analysis are given in Table 3 and are shown in Fig. 4 by circles. The phase shift in Fig. 4 started from 180° because, as shown in [48], the *S* wave should have a forbidden state. Here and further, when we use excitation functions, the cross sections obtained on the basis of the found phase shifts absolutely lay in the limit of available experimental error.

Table 3. $^2S_{1/2}$ scattering phase shift that we obtained from [39].

| $E_p$, MeV | Phase shift, deg |
|---|---|
| 0.3855 | 179.59 |
| 0.4871 | 179.74 |
| 0.6162 | 179.65 |
| 0.6631 | 178.11 |
| 0.7162 | 177.96 |
| 0.759 | 176.46 |
| 0.8108 | 174.55 |
| 0.8612 | 173.37 |
| 0.9058 | 174.19 |
| 0.979 | 172.17 |
| 1.1063 | 169.82 |
| 1.2508 | 166.43 |
| 1.3704 | 163.61 |
| 1.5898 | 158.42 |
| 1.7903 | 155.78 |
| 1.9909 | 151.09 |

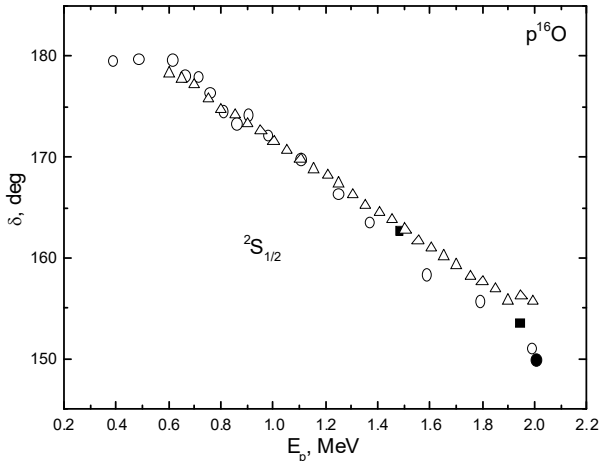

Fig. 4. Phase shifts of p$^{16}$O elastic scattering that we obtained from excitation functions from [39] – open circles and from [37] – open triangles. A comparison to the results of phase shift analysis of [28] (dot) and [33] (squares) at energies above 1.5 MeV is given.

The comparison of results of the phase analysis of [28] – the dot at 2 MeV and [33] – squares at about 1.5 MeV and 2 MeV are given in Fig. 4. Among the new results given in Fig. 4 by circles one can clearly see the form of the $^2S_{1/2}$ scattering phase shift at the lowest energies, which plays a major role in the consideration of thermonuclear processes of radiative capture at astrophysical energies. At an energy of 0.6 MeV and less, this phase shift is almost equal to 180°. At energies above 1.5 MeV there is good agreement with previous results of phase shift analyses [28,33]. The difference between present and previous results on scattering phase shifts obtained in the 1960s does not exceed 1°–2°. Here one can see that the differential cross section measurements in [39] were made in the mid-1970s, and there is not much difference from the data of [28,33]. However, in other studies, another value of the $\hbar^2/m_0$ constant could be used; this fact can explain such a difference in the phase shifts. The partial $\chi^2$ value with the experimental errors of the cross sections in the excitation functions given in [39] does not exceed $10^{-5}$. As a result, one $S_{1/2}$ scattering phase shift is completely unique. For all other results of phase shift analyses made for the excitation functions given further, the partial $\chi^2$ value is approximately the same.

Now consider data on the excitation functions from [37] at energies 0.6–2.0 MeV and scattering angle 160° in l.s. or 161.2° in c.m. Cross sections given in [37] are given in l.s.; we converted them into in c.m. They are given by points with 5% errors mentioned in [37]. To convert the cross sections we used an expression

$$\sigma_{cm} = \left| \frac{1 + \gamma \cos(\theta_{cm})}{1 + \gamma^2 + 2\gamma \cos(\theta_{cm})} \right|^{1.5} \sigma_{lab},$$

where $\gamma = m_1/m_2$, $m_1$ is incident particle and $m_2$ – mass of target nucleus. We used whole values of particle masses in these conversions.

As seen from Fig. 4, where the triangles are the scattering phase shifts obtained using data from [37], the results of our phase shift analysis are in good agreement with the previous phase shift extractions at energies of up to 2 MeV. The exception is the last two points in the scattering phase shifts, where the excitation functions [37] for these energies are also slightly different to other results.

Furthermore [49], results for the differential cross sections at energies from 0.8–2.5 MeV and angle 170° in l.s. or 170.6° in c.m. with 4% error were obtained. However, the phase shift analysis of these data, as far as we know, has not yet been made. The results for excitation functions correspond to scattering phase shifts obtained in our analysis are shown in Fig. 5 by down triangles. It is clearly seen from Fig. 5 that the results of data from [39] and [49] coincide, although the measurements of the excitation functions were made with an interval of 10 years. The open squares in Fig. 5 are the results of the phase shift analysis of angular distributions taking into account only the $^2S_{1/2}$ phase shifts, given in [49] at energies approximately from 1.8–2.4 MeV.

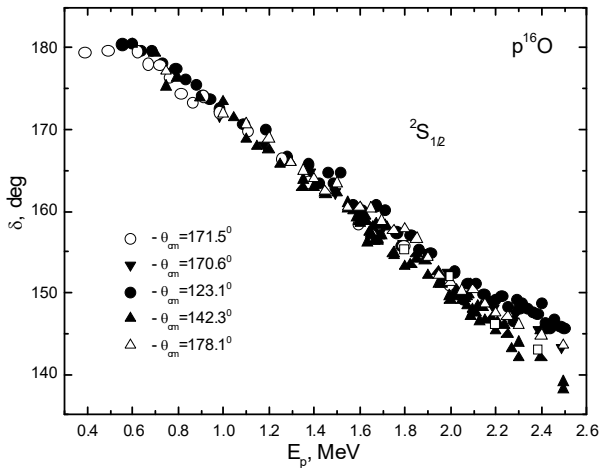

Fig. 5. Triangles - phases of p$^{16}$O elastic scattering that we obtained from the excitation functions of [49]. Circles - phase shifts from Fig. 4 obtained using data from [39]. Open squares represent the results of our phase shift analysis of angular distributions of [49].

Excitation functions for angles of 90° and 120° in l.s. or 93.6° and 123.1° in c.m. in the proton energy range of 0.5–3.5 MeV with 5% experimental errors were measured in [50]. Here we used the results for the second angle at energy 2.5 MeV; the solid curve shows the agreement between the experimental cross sections and those calculated with the obtained phase shifts, which are shown in Fig. 5 by the black points. The circles in Fig. 5 are our results on the basis of data from [39] at 171.5° in c.m., which, as seen in Fig. 4, are in acceptable agreement with the earlier phase shift analysis results of [28,33]. Measurements of [50] were made in the late 1990s and data in [28,33] were published in the 1960s. However, the results of the phase shift analysis made on these data in the energy range of 0.4–2.5 MeV are in quite acceptable agreement with each other, as shown in Figs. 4 and 5.

Let us now consider newer experimental data on excitation functions [38] and their phase shift analysis. Excitation functions at energy range from 0.6–2.5 MeV at angles 140° and 178° in l.s. or 142.3° and 178.1° in c.m. were measured in [38]. Fig. 5 shows the results of our phase shift analysis obtained from these excitation functions. One can see from these results the form of the $^2S_{1/2}$ scattering phase shift at lowest energies, which exceeds 180° by 1°–2°. Let us recall that, since the phase shift analysis is made by a single point in the cross sections, i.e., one value of the cross section at a given energy, the $^2S_{1/2}$ scattering phase shift is completely unique. Therefore, such an excess of 180° in scattering phase shifts may indicate a real error in the determination of these phase shifts of the considered experimental data.

One can see in this figure that the agreement between scattering phase shifts higher than 0.8 MeV, obtained in 1975 (see [39]) and more recent data [38] published in 2002 is better than in previous cases. For previous results obtained from the data of [39], and from [49,50], the greater difference between the phase shifts was shown. However, at energies higher than 2.2 MeV the results of our analysis show a significant difference in the scattering phase shifts obtained for these two angles, which reach 4°–5° at an energy of 2.5 MeV.

In conclusion let us consider the results of the phase shift analysis of the current measurements of differential cross sections in the excitation functions and angular distributions of p$^{16}$O elastic scattering given in Tables 1 and 2. The cross sections calculated with the obtained scattering phase shifts and the phase shifts are given in Fig. 6 by black triangles. Points, circles and squares in the same figure show the comparison of our phase shifts obtained using various experimental data. It is seen that phase shifts lower than 0.8 MeV, obtained using our data, are located slightly below the results of our analysis of data from [39], and given by circles – this difference equals 2°–3°. Starting from 0.8 MeV to 1.0 MeV, the results for our data coincide with the phase shifts for the data from [39] to an accuracy of about 1°.

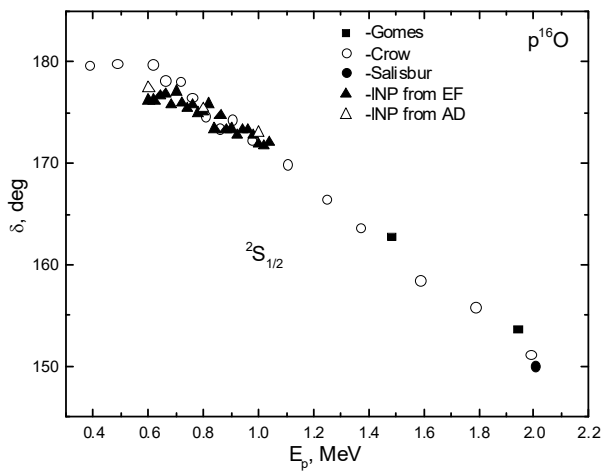 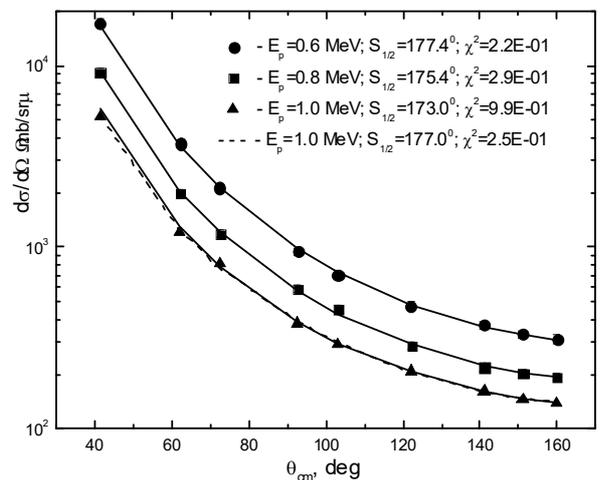

Fig. 6. Phase shifts of p$^{16}$O elastic scattering we obtained from the differential cross sections in the current work are triangles. Other denotations = data of [28,33,39].

Fig. 7. Angular distributions of p$^{16}$O elastic scattering measured at proton energies of 0.6, 0.8 and 1.0 MeV. Solid and dashed curves - cross sections calculated with scattering phase shifts obtained in our phase shift analysis.

Furthermore, the phase shifts obtained from the measured angular distributions are presented in Fig. 6 by open triangles, and the cross sections' description quality and the value of $\chi^2$ at three energies are given in Fig. 7. It is seen from Fig. 7 that the $\chi^2$ value for $E = 1.0$ is quite large; almost equal to 1. Here components of the cross sections expansion by partial waves that are higher than $^2S$ wave can contribute noticeably. Taking into account the contribution of $^2P$ waves with 20 iterations [4] leads to the following values of the scattering phase shifts: $S = 172.4°$, $P_{1/2} = 2.4°$, $P_{3/2} = 2.4°$ at $\chi^2 = 0.68$. Taking into account the contribution of the $P$ and $D$ waves with the same number of iterations leads to the following scattering phase shifts: $S = 177.0°$, $P_{1/2} = -8.8°$, $P_{3/2} = 10.9°$, $D_{3/2} = 1.7°$, $D_{5/2} = 1.3°$ at $\chi^2 = 0.25$. The results of cross sections calculations are given in Fig. 7 by the dashed curve. The improvement in the cross sections' description is observed only in the forward angular range up to about 70°–80°.

These phase shifts are in agreement with the phase shifts obtained using excitation functions from Table 1 at energies of less than 0.8 MeV are also located lower than the results of the phase shift analysis obtained using data from [39]. However, the accuracy of each phase shift analysis estimated by us is 2°–3°, so the observed difference in the results is within such errors.

## 5. Conclusions

We have obtained new results for the differential cross sections and phase shifts of p$^{16}$O elastic scattering in the description of these and other data on the excitation functions from several works for different scattering angles in the energy range of 0.4–2.5 MeV. Quite a good agreement of all the obtained results (Fig. 5) with each other and the phase shift analysis at energy up to about 2.0 MeV made previously is observed. For example, the difference of the phase shifts at the energy of 2.0 MeV obtained on the basis of different data, starting from [28] (carried out in 1962) and our modern measurements (2015), equals about 3° at the phase shifts values of about 150°–155°. Consequently, this difference equals approximately 2% at the experimental errors of the elastic cross sections of 5–10%.

The results of the phase shift analysis carried out, i.e., phase shifts of p$^{16}$O elastic scattering and the data on resonances of $^{17}$F [32], will allow in the future to parameterize interclusteral interaction potentials for scattering processes in a non-resonant $^2S_{1/2}$ wave. These potentials, in turn, may further be used in calculations of various astrophysical problems, such as radiative capture of particles on light nuclei, and in this case, proton capture on $^{16}$O. Some have already been considered and the results have been given, for example, in books [2,16] or reviews [5,9–13].

## Acknowledgments

The work was performed under grant No. 0073/PCF-15-MES "The astrophysical study of stellar and planetary systems" No.10 "Studying of the thermonuclear processes in the Universe" of the Ministry of Education and Science of the Republic of Kazakhstan.